\documentclass[twocolumn,showpacs,preprintnumbers,pra]{revtex4}
\usepackage{amssymb}
\usepackage{amsmath}
\usepackage{graphicx}
\usepackage{dcolumn}
\usepackage{bm}

\begin{document}

\title{Entangling Power in the Deterministic Quantum Computation with One
Qubit }
\author{Chang-shui Yu$^1$}
\email{quaninformation@sina.com; ycs@dlut.edu.cn}
\author{X. X. Yi$^1$}
\author{He-shan Song$^1$}
\author{Heng Fan$^2$}
\affiliation{$^1$School of Physics and Optoelectronic Technology, Dalian University of
Technology, Dalian 116024, P. R. China\\
$^{2}$Beijing National Laboratory for Condensed Matter Physics, Institute of
Physics, Chinese Academy of Sciences, Beijing 100190, China\\
}
\date{\today }

\begin{abstract}
The deterministic quantum computing with one qubit (DQC1) is a mixed-state
quantum computation algorithm that evaluates the normalized trace of a
unitary matrix and is more powerful than the classical counterpart. We find
that the normalized trace of the unitary matrix can be directly described by
the entangling power of the quantum circuit of the DQC1, so the nontrivial
DQC1 is always accompanied with the non-vanishing entangling power. In
addition, it is shown that the entangling power also determines the
intrinsic complexity of this quantum computation algorithm, i.e., the larger
entangling power corresponds to higher complexity. Besides, it is also shown
that the non-vanishing entangling power does always exist in other similar
tasks of DQC1.
\end{abstract}

\maketitle

\section{Introduction}

Quantum entanglement is employed in most of quantum information processing
tasks (QIPTs) including the quantum algorithms and the quantum
communications [1]. There is no doubt that quantum entanglement is an
important physical resource in quantum information processing. However,
quantum entanglement cannot be competent for all the QIPTs [2-5]. Strong
evidence has shown that some QIPTs display the quantum advantage, but there
does not exist any entanglement in the tasks [6,7], which is also verified
in experiment [8]. One such remarkable evidence is the scheme of the
deterministic quantum computing with one qubit (DQC1) which accomplishes to
evaluate the normalized trace of a unitary matrix by only measuring a
control qubit, irrespective of the complexity of the unitary matrix of
interests [9]. But DQC1 can not be performed effectively by only using
classical computation [9,10]. So what quantum property leads to the quantum
advantage in the DQC1?

Quantum discord was introduced to effectively distinguish the quantum from
the classical correlations [11-13]. It is shown that quantum correlation
`includes' quantum entanglement but beyond it due to the presence in
separable states [14-16]. In recent years, quantum discord has attracted
increasing interests especially in the areas such as the evolution in
quantum dynamical system [17-19], operational interpretations [20-26] and
the quantification [27,28] and the experiments \cite{experim1,experim2}. In
particular, it is found that almost every unitary matrix (random unitary) in
the DQC1 will lead to the occurrence of quantum discord in the output state,
so it is conjectured that quantum discord could be the quantum nature of the
DQC1 [10]. However, after all, there exist some general unitary matrices
(for example, Hermitian unitary matrices) that will arrive at the output
states without any quantum discord. In this sense, it seems that quantum
discord should not be the source of the quantum advantage of the DQC1
either, as is first suspected in Ref. \cite{[29]}. Thus what on earth is the
quantum nature of DQC1 is still left open.

In this paper, we find that the trace calculation and the complexity in the
DQC1 can be directly related to the entangling power which is defined by the
maximal average on the ability to entangle a qubit with a pure state in a
given ensemble. We find that the entangling power of the DQC1 circuit can be
directly written as the normalized trace of the unitary transformation to be
measured. Based on this result, we find not only that in the nontrivial DQC1
is there the entangling power, but also that the intrinsic complexity (which
will be given at the end) of the evaluation of the normalized trace is
determined by the entangling power. In this sense, we say that the
entangling power could be used to signal the quantum nature of the DQC1. As
a supplement, we also consider the other QIPTs using the similar DQC1
circuit. One will find that the entangling power is always present if the
information of the system can be extracted by the control qubit.

\begin{figure}[tbp]
\includegraphics[width=0.6\columnwidth]{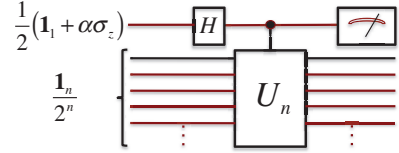}
\caption{The generalized DQC1 circuit with arbitrary polarization ($0<%
\protect\alpha \leq 1$).}
\label{1}
\end{figure}

\section{\protect\bigskip The entangling power with the nomalized trace}

We begin with a brief introduction of the generalized DQC1 described by the
quantum circuit given in Fig. 1 [9]. The initial state can be written as 
\begin{equation}
\rho _{0}=\rho _{c}^{\alpha }\otimes \frac{\mathbf{1}_{n}}{2^{n}},
\end{equation}%
where $\rho _{c}^{\alpha }=\frac{1}{2}\left( \mathbf{1}_{1}+\alpha \sigma
_{z}\right) $, $\sigma _{x,y,z}$ are Pauli matrices, the subscript `c'
denotes the control qubit, the superscript $\alpha $ means that the density
matrix depends on the parameter $\alpha $ and $\mathbf{1}_{n}$ means the
identity of $n$ qubits. It is noted that in the standard DQC1 [7], the
initial control qubit is given by $\left\vert 0\right\rangle $ . Through the
quantum circuit, the state given in Eq. (1) will be transformed into the
final state $\rho _{n+1}$, 
\begin{eqnarray}
\rho _{n+1} &=&\frac{1}{2^{n}}\left( \left\vert 0\right\rangle \left\langle
0\right\vert \otimes \mathbf{1}_{n}+\left\vert 1\right\rangle \left\langle
1\right\vert \otimes \mathbf{1}_{n}\right.   \notag \\
&&+\left. \alpha \left\vert 0\right\rangle \left\langle 1\right\vert \otimes
U_{n}^{\dag }+\alpha \left\vert 1\right\rangle \left\langle 0\right\vert
\otimes U_{n}\right) .
\end{eqnarray}%
Thus if we measure the control qubit in the basis of $\sigma _{x}$ and $%
\sigma _{y}$, respectively, one can obtain the corresponding expectations as 
$\frac{1}{2^{n}}$Re(Tr$U_{n}$) and $-\frac{1}{2^{n}}$Im(Tr$U_{n}$). In this
way, the normalized trace of the unitary matrix $U_{n}$ is obtained only by
the measurements on a single qubit, irrespective of the complexity of the
unitary matrix. This shows the quantum advantage of the DQC1 in the
reduction of the computational complexity.

In this task, the initial state $\rho _{0}$ is obviously mixed, so in
practical scenario, it has to be prepared by one of its pure-state
realizations $\left\{ p_{i},\left\vert \varrho _{i}\right\rangle \right\} $
such that $\rho _{0}=\sum p_{i}\left\vert \varrho _{i}\right\rangle
\left\langle \varrho _{i}\right\vert $, $\sum p_{i}=1,$where $\left\vert
\varrho _{i}\right\rangle $ is normalized but not necessarily orthogonal. An
intuitive observation shows that the DQC1 circuit can lead to the entangled
final state of $\left\vert \varrho _{i}\right\rangle $ if any information on 
$U_{n}$ is unknown. Therefore, it is not difficult to ignite the light to
relate this kind of entanglement to the mechanism of the quantum advantage
of the DQC1. To do so, we would like to introduce the variational concept of
entangling power. It is initially defined for a unitary transformation by
measuring the average entanglement produced by this unitary transformation
on the separable state subject to some kind of distributions \cite{[30]}.
However, in some special QIPTs, not all the quantum separable states are
covered, so it is necessary to give an explicit definition well-suited to
the given QIPT. So in the DQC1, we would like to consider the entangling
power of the controlled unitary transformation $\mathbf{1}_{n-1}\oplus U_{n}$
with the assistance of the Hadamard gate $H$ subject to the initial ensemble 
$\frac{\mathbf{1}_{n}}{2^{n}}$. In other words, we will quantify the ability
of the whole DQC1 circuit to entangle the control qubit with the $n$-qubit
pure state selected from the initial ensemble $\frac{\mathbf{1}_{n}}{2^{n}}$%
. With this aim, we have the following rigid definition.

\textbf{Definition}.- The entangling power of the DQC1 circuit is given by%
\begin{equation}
E_{p}\left( \tilde{U}_{n}\right) =\max_{\left\{ q_{i},\left\vert \varphi
_{i}\right\rangle \right\} }\sum\limits_{i}q_{i}E\left[ \tilde{U}_{n}\left(
\rho _{c}\otimes \left\vert \varphi _{i}\right\rangle \left\langle \varphi
_{i}\right\vert \right) \tilde{U}_{n}^{\dag } \right]
\end{equation}%
with $\tilde{U}_{n} =\left(\mathbf{1}_{n-1}\oplus U_{n}\right)\times\left (
H\otimes\mathbf{1}_{n}\right)$, $\frac{\mathbf{1}_{n}}{2^{n}}=\sum
q_{i}\left\vert \varphi _{i}\right\rangle \left\langle \varphi
_{i}\right\vert $, where $E\left[ \cdot \right] $ represents any a good
entanglement measure [1].

Here we let $E\left[ \cdot \right] =\sqrt{2\left( 1-\text{Tr}\rho
_{r}^{2}\right) }$ with $\rho _{r}$ the reduced density matrix of the state
taken into account [1]. The maximum is taken due to the non-uniqueness of
the realization of $\frac{\mathbf{1}_{n}}{2^{n}}$. In addition, $\rho _{c}$
is not limited to the pure state, which is different from the original
definition of the entangling power besides the limited ensemble. Next, we
will give our main results on $E_{p}\left( \tilde{U}_{n}\right) $ by two
theorems.

\textbf{Theorem 1.}-The entangling power, defined in Eq. (3) for the
standard DQC1 circuit corresponding to $\rho _{c}^{1}=\left\vert
0\right\rangle \left\langle 0\right\vert ,$i.e., $\alpha =1$, is given by 
\begin{equation}
E_{p}^{1}\left( \tilde{U}_{n}\right) =\sqrt{1-\left\vert \frac{\text{Tr}U_{n}%
}{2^{n}}\right\vert ^{2}}.
\end{equation}%
\textbf{Proof.} Substitute $\rho _{c}^{1}=\left\vert 0\right\rangle
\left\langle 0\right\vert $ and any an $n$-partite pure state $\left\vert
\varphi _{i}\right\rangle \ $chosen from the ensemble $\frac{\mathbf{1}_{n}}{%
2^{n}}=\sum q_{i}\left\vert \varphi _{i}\right\rangle \left\langle \varphi
_{i}\right\vert $ into the DQC1 circuit sketched in Fig. 1, the final state
after these substitutions can be written as

\begin{equation}
\left\vert \chi \right\rangle _{n+1}=\frac{1}{\sqrt{2}}\left( \left\vert
0\right\rangle \left\vert \varphi _{i}\right\rangle +\left\vert
1\right\rangle U_{n}\left\vert \varphi _{i}\right\rangle \right) .
\end{equation}%
The reduced density matrix by tracing out the control qubit is given by%
\begin{equation}
\rho _{r}^{i}=\frac{1}{2}\left\vert \varphi _{i}\right\rangle \left\langle
\varphi _{i}\right\vert +U_{n}\left\vert \varphi _{i}\right\rangle
\left\langle \varphi _{i}\right\vert U_{n}^{\dag }.
\end{equation}%
So the entangling power can be expressed as%
\begin{eqnarray}
E_{p}^{1}\left( \tilde{U}_{n}\right) &=&\max_{\left\{ q_{i},\left\vert
\varphi _{i}\right\rangle \right\} }\sum\limits_{i}q_{i}\sqrt{2\left(
1-Tr\left( \rho _{r}^{i}\right) ^{2}\right) }  \notag \\
&=&\max_{\left\{ q_{i},\left\vert \varphi _{i}\right\rangle \right\}
}\sum\limits_{i}q_{i}\sqrt{1-\left\vert \left\langle \varphi _{i}\right\vert
U_{n}\left\vert \varphi _{i}\right\rangle \right\vert ^{2}} \\
&\leq &\sqrt{1-\left\vert \sum\limits_{i}q_{i}\left\langle \varphi
_{i}\right\vert U_{n}\left\vert \varphi _{i}\right\rangle \right\vert ^{2}} 
\notag \\
&=&\sqrt{1-\left\vert \frac{\text{Tr}U_{n}}{2^{n}}\right\vert ^{2}}.
\end{eqnarray}%
The inequality comes from the concave property of the entanglement measure $E%
\left[ \cdot \right] $ and the maximum is attained by the realization $\frac{%
\mathbf{1}_{n}}{2^{n}}=\sum \tilde{q}_{i}\left\vert \tilde{\varphi}%
_{i}\right\rangle \left\langle \tilde{\varphi}_{i}\right\vert $ where $%
\tilde{q}_{i}=\frac{1}{2^{n}}$ and $\left\vert \tilde{\varphi}%
_{j}\right\rangle =\sum\limits_{k}e^{i\frac{2jk\pi }{2^{n}}}\left\vert
\upsilon _{k}\right\rangle $ with $\left\vert \upsilon _{k}\right\rangle $
the eigenvectors of $U_{n}$. The proof is completed.\hfill$\blacksquare$

It is quite interesting that the entangling power for the standard DQC1 is
directly described by the normalized trace of the measured $U_{n}$. So long
as Tr$U_{n}\neq 2^{n}$ which implies $U_{n}\neq \mathbf{1}_{n}e^{i\theta }$,
the entangling power will not vanish. This means that the DQC1 will
demonstrate the quantum advantage. Otherwise, the entangling power will
vanish for $U_{n}=\mathbf{1}_{n}e^{i\theta }$, but the unitary matrix $U_{n}$
in this case will be easily evaluated in the classical computation. So it is
a trivial case.

\textbf{Theorem 2.}-The entangling power for the generalized DQC1 circuit
corresponding to $\rho _{c}^{\alpha }=\frac{1}{2}\left( \mathbf{1}%
_{1}+\alpha \sigma _{z}\right) $, i.e., $0<\alpha <1$, is given by 
\begin{equation}
E_{p}^{\alpha }\left( \tilde{U}_{n}\right) =\alpha E_{p}^{1}\left( \tilde{U}%
_{n}\right) =\alpha \sqrt{1-\left\vert \frac{\text{Tr}U_{n}}{2^{n}}%
\right\vert ^{2}}.
\end{equation}%
\textbf{Proof. }If $0<\alpha <1$, the control qubit is obviously a mixed
state. Based on the definition of entangling power given in Eq. (3), we can
write 
\begin{equation}
E_{p}^{\alpha }\left( \tilde{U}_{n}\right) =\max_{\left\{ q_{i},\left\vert
\varphi _{i}\right\rangle \right\} }\sum\limits_{i}q_{i}E\left[ \varrho _{i}%
\right] ,
\end{equation}%
with 
\begin{equation}
\varrho _{i}=\left( \mathbf{1}_{n-1}\oplus U_{n}\right) \left( H\rho
_{c}^{\alpha }H^{\dag }\otimes \left\vert \varphi _{i}\right\rangle
\left\langle \varphi _{i}\right\vert \right) \left( \mathbf{1}_{n-1}\oplus
U_{n}\right) .
\end{equation}%
Since $E\left[ \varrho _{i}\right] $ denotes the entanglement measure of the
state $\varrho _{i}$, we have 
\begin{equation}
E\left[ \varrho _{i}\right] =\min_{\left\{ r_{j},\left\vert \gamma
_{j}\right\rangle \right\} }\sum_{j}r_{j}E_{i}\left[ \left\vert \gamma
_{j}\right\rangle \right] ,
\end{equation}%
with $\varrho _{i}=\sum_{j}r_{j}\left\vert \gamma _{j}\right\rangle
\left\langle \gamma _{j}\right\vert $ and the subscript $i$ corresponding to 
$\varrho _{i}$. Thus the entangling power can be rewritten as%
\begin{eqnarray}
E_{p}^{\alpha }\left( \tilde{U}_{n}\right) &=&\max_{\left\{ q_{i},\left\vert
\varphi _{i}\right\rangle \right\} }\sum\limits_{i}q_{i}\min_{\left\{
r_{j},\left\vert \gamma _{j}\right\rangle \right\} }\sum_{j}r_{j}E_{i}\left[
\left\vert \gamma _{j}\right\rangle \right]  \notag \\
&=&\min_{\left\{ r_{j},\left\vert \gamma _{j}\right\rangle \right\}
}\max_{\left\{ q_{i},\left\vert \varphi _{i}\right\rangle \right\}
}\sum\limits_{ij}q_{i}r_{j}E_{i}\left[ \left\vert \gamma _{j}\right\rangle %
\right] ,
\end{eqnarray}%
where the exchange of the maximum and minimum is attributed to the
independence of realizations $\left\{ r_{j},\left\vert \gamma
_{j}\right\rangle \right\} $ and $\left\{ q_{i},\left\vert \varphi
_{i}\right\rangle \right\} $. Consider the eigendecomposition of $\varrho
_{i}$: $\varrho _{i}=\Phi _{i}M\Phi _{i}^{\dag }$, one can obtain that 
\begin{eqnarray}
\Phi _{i} &=&\left( \mathbf{1}_{n-1}\oplus U_{n}\right) \left( H\otimes
\left\vert \varphi _{i}\right\rangle \right) , \\
M &=&\left( 
\begin{array}{cc}
\frac{1+\alpha }{2} & 0 \\ 
0 & \frac{1-\alpha }{2}%
\end{array}%
\right) .
\end{eqnarray}%
Thus any decomposition of $\varrho _{i}$ can be given in terms of the
eigendecomposition characterized by Eqs. (14) and (15). Let $\varrho
_{i}=\sum_{j}r_{j}\left\vert \gamma _{j}\right\rangle \left\langle \gamma
_{j}\right\vert $ be one decomposition with the form of matrix given by $%
\varrho _{i}=\Psi W\Psi ^{\dagger }$ where the columns of $\Psi $ correspond
to $\left\vert \gamma _{j}\right\rangle $ and the diagonal entries of the
diagonal matrix $W$ correspond to $r_{j}$. Therefore, we have $\Psi W^{\frac{%
1}{2}}=\Phi _{i}M^{\frac{1}{2}}T$ where $T$ denotes a right unitary matrix
with $TT^{\dag }=\mathbf{1}_{1}$. So $\left\vert \gamma _{j}\right\rangle $
can be given by the eigenvectors as 
\begin{equation}
\left\vert \gamma _{j}\right\rangle =\frac{1}{\sqrt{r_{j}}}\left(
x_{j}\left\vert 0\right\rangle \left\vert \varphi _{i}\right\rangle
+y_{j}\left\vert 1\right\rangle U_{n}\left\vert \varphi _{i}\right\rangle
\right) ,
\end{equation}%
where 
\begin{eqnarray}
x_{j} &=&\frac{1}{\sqrt{2}}\left( T_{1j}\sqrt{\frac{1+\alpha }{2}}+T_{2j}%
\sqrt{\frac{1-\alpha }{2}}\right) , \\
y_{j} &=&\frac{1}{\sqrt{2}}\left( T_{1j}\sqrt{\frac{1+\alpha }{2}}-T_{2j}%
\sqrt{\frac{1-\alpha }{2}}\right)
\end{eqnarray}%
and $r_{j}=\left\vert x_{j}\right\vert ^{2}+\left\vert y_{j}\right\vert ^{2}$
with $\left\vert T_{1j}\right\vert ^{2}+\left\vert T_{2j}\right\vert ^{2}$ $%
\leq 1$ and 
\begin{equation}
\sum_{j}\left\vert T_{1j}\right\vert ^{2}=\sum_{j}\left\vert
T_{2j}\right\vert ^{2}=1.
\end{equation}%
Substitute Eq. (16) into Eq. (13), we can arrive at

\begin{eqnarray}
E_{p}^{\alpha }\left( \tilde{U}_{n}\right) &=&\min_{\left\{ r_{j},\left\vert
\gamma _{j}\right\rangle \right\} }\max_{\left\{ q_{i},\left\vert \varphi
_{i}\right\rangle \right\} }\sum\limits_{i,j}\sqrt{2}q_{i}\left(
r_{j}^{2}\right.  \notag \\
&&-\left\vert x_{j}\right\vert ^{4}-\left\vert y_{j}\right\vert ^{4}\left.
-2\left\vert x_{j}\right\vert ^{2}\left\vert y_{j}\right\vert ^{2}\left\vert
\left\langle \varphi _{i}\right\vert U_{n}\left\vert \varphi
_{i}\right\rangle \right\vert ^{2}\right) ^{\frac{1}{2}}  \notag \\
&=&\min_{\left\{ r_{j},\left\vert \gamma _{j}\right\rangle \right\}
}\sum\limits_{j}2\left\vert x_{j}\right\vert \left\vert y_{j}\right\vert 
\notag \\
&&\times \max_{\left\{ q_{i},\left\vert \varphi _{i}\right\rangle \right\}
}\sum\limits_{i}q_{i}\sqrt{1-\left\vert \left\langle \varphi _{i}\right\vert
U_{n}\left\vert \varphi _{i}\right\rangle \right\vert ^{2}}.
\end{eqnarray}%
Based on Eq. (7) (or Theorem 1), we have 
\begin{equation}
E_{p}^{\alpha }\left( \tilde{U}_{n}\right) =\min_{\left\{ r_{j},\left\vert
\gamma _{j}\right\rangle \right\} }\sum\limits_{j}2\left\vert
x_{j}\right\vert \left\vert y_{j}\right\vert E_{p}^{1}\left( \tilde{U}%
_{n}\right) .
\end{equation}%
Insert Eqs. (17) and (18) into Eq. (21) and use Eq. (19), it follows%
\begin{eqnarray}
&&E_{p}^{\alpha }\left( \tilde{U}_{n}\right)  \notag \\
&=&\min_{\left\{ r_{j},\left\vert \gamma _{j}\right\rangle \right\} }\frac{1%
}{2}\sum\limits_{j}\left\vert \left( 1+\alpha \right) T_{1j}^{2}-\left(
1-\alpha \right) T_{2j}^{2}\right\vert E_{p}^{1}\left( \tilde{U}_{n}\right) 
\notag \\
&\geq &\frac{1}{2}\sum\limits_{j}\left[ \left( 1+\alpha \right) \left\vert
T_{1j}\right\vert ^{2}-\left( 1-\alpha \right) \left\vert T_{2j}\right\vert
^{2}\right] E_{p}^{1}\left( \tilde{U}_{n}\right)  \notag \\
&=&\alpha E_{p}^{1}\left( \tilde{U}_{n}\right) =\alpha \sqrt{1-\left\vert 
\frac{\text{Tr}U_{n}}{2^{n}}\right\vert ^{2}}
\end{eqnarray}%
where the minimum is achieved when both $T_{1j}$ and $T_{2j}$ are real or
imaginary for all $j$. The proof is completed.\hfill$\blacksquare$

From the above two theorems, one can find the result given in Theorem 2 can
be reduced to Theorem 1 for $\alpha =1$. That is, the entangling power $%
E_{p}^{\alpha }\left( \tilde{U}_{n}\right) $ pertains to all the cases of $%
\alpha $. In addition, the entangling power of the generalized DQC1
including the standard case is directly described by the normalized trace of 
$U_{n}$ to be measured. As is the same as the analysis of the standard DQC1,
if $U_{n}\neq \mathbf{1}_{n}e^{i\theta }$ which is a trivial case, the DQC1
will demonstrate the quantum advantage of the quantum computation, the
entangling power will not vanish. In addition, it can be easily found that
the entangling power will increase as $\alpha $ increasing. When $\alpha =0$%
, this means that the control qubit is an identity which will extract
nothing about the measured $U_{n}$. In this case, the entangling power
vanishes, which is consistent with our expectation. 

\section{\protect\bigskip The complexity with the entangling power}

In fact, our entangling power is also closely related to the complexity of
the DQC1. The complexity of the DQC1 can be given by 
\begin{equation}
O(U_{n})=nL(\varepsilon ),
\end{equation}%
where $L(\varepsilon )$ is the measurement complexity which describes how
many rounds of measurements are necessary to be operated on the control
qubit for a given standard deviation $\varepsilon $, and $n$ is the input
complexity which denotes the number of qubits needed to be input into the
DQC1 circuit. In usual analysis of the complexity of DQC1, only the
measurement complexity $L(\varepsilon )$ is considered. So it is stated that
the complexity $L(\varepsilon )$ only depends on the accuracy $\varepsilon $
that we expect instead of the scale of the measured $U_{n}$, because%
\begin{equation}
L(\varepsilon )\thicksim \ln (1/P_{e})/\left( \alpha ^{2}\varepsilon
^{2}\right) ,
\end{equation}%
where $P_{e}$ is the probability that the estimate is farther from the true
value than $\varepsilon $ [10]. When we consider the practical experiment,
the standard deviation should not exceed the true value of the measured
quantity. In this sense, the complexity will also depend on the true value
of the measured observables to different extents. Thus, instead of the
standard deviation $\varepsilon $, it would be reasonable to describe the
accuracy in terms of the relative error defined by 
\begin{equation}
\epsilon =\left\vert \frac{\varepsilon }{X}\right\vert \times 100\%,
\end{equation}%
with $X$ the true value of some measured quantities. In the DQC1, $\sigma
_{x}$ and $\sigma _{y}$ will be measured on the control qubit corresponding
to the normalized trace of $U_{n}$ (actually to the real and imaginary
parts, respectively). Let our finally expecting relative errors be $\epsilon
\geq \max \{\epsilon (\sigma _{x}),\epsilon (\sigma _{y})\}$ with $\epsilon
(\sigma _{x})$ and $\epsilon (\sigma _{y})$ denoting the expecting relative
errors for the measurements $\sigma _{x}$ and $\sigma _{y}$ such that 
\begin{equation}
\frac{\ln \left[ 1/P_{e}(\sigma _{x})\right] }{\left\vert \epsilon (\sigma
_{x})\text{Re}\left( \frac{\text{Tr}U_{n}}{2^{n}}\right) \right\vert ^{2}}=%
\frac{\ln \left[ 1/P_{e}(\sigma _{y})\right] }{\left\vert \epsilon (\sigma
_{y})\text{Im}\left( \frac{\text{Tr}U_{n}}{2^{n}}\right) \right\vert ^{2}}%
\thicksim \alpha ^{2}L.
\end{equation}%
Substitute Eqs. (24-26) into Eq. (9), the entangling power can be written as%
\begin{equation}
E_{p}^{\alpha }\left( \tilde{U}_{n}\right) \thicksim \sqrt{\alpha ^{2}-\frac{%
M}{L}},
\end{equation}%
with $M=\frac{\ln \left[ 1/P_{e}(\sigma _{x})\right] }{\left\vert \epsilon
(\sigma _{x})\right\vert ^{2}}+\frac{\ln \left[ 1/P_{e}(\sigma _{y})\right] 
}{\left\vert \epsilon (\sigma _{y})\right\vert ^{2}}$. Thus for a given $M$,
the measurement complexity $L$ is directly determined by the entangling
power $E_{p}^{\alpha }\left( \tilde{U}_{n}\right) $. Since $M$ is determined
by the expecting errors and is independent of the scale of the measured $%
U_{n}$, we think that the complexity $L$ with the fixed $M$ is the intrinsic
complexity. The larger entangling power means the larger intrinsic complexly 
$L$. Thus one can find that the intrinsic complexity is directly determined
by the entangling power. 

\section{\protect\bigskip DQC1-like circuits in other tasks}
We would like to generalize the current DQC1 circuit to
general QIPTs, by which one will find that the non-trivial tasks with DQC1
circuit are indeed accompanied with non-vanishing entangling power. Suppose
this circuit is used to extract the information of some state $\rho _{n}$
instead of $\mathbf{1}_{n}$ and the control qubit is the general quantum
state%
\begin{equation}
\rho _{c}=\frac{1}{2}\left( \mathbf{1}_{1}+\mathbf{P}\cdot \mathbf{\sigma}
\right)
\end{equation}
with $\mathbf{P}$ the polarization vector and $\mathbf{\sigma }$ the
corresponding vector of Pauli matrices. The final state $\rho _{f}$ of $\rho
_{c}$ via the circuit can be written as $\rho _{f}=$ Tr$_{n}\tilde{U}%
_{n}\left( H\rho _{c}H^{\dag }\otimes \rho _{n}\right) \tilde{U}_{n}^{\dag }$%
. The linear entropy of $\rho _{f}$ is given by 
\begin{eqnarray}
L(\rho _{f}) &=&1-Tr\rho _{f}^{2}  \notag \\
&=&\frac{1}{2}(1-P_{1}^{2}-(P_{2}^{2}+P_{3}^{2})\left\vert \text{Tr}U\rho
_{n}\right\vert ^{2}).
\end{eqnarray}%
In order to accomplish this extraction of information, $L(\rho _{f})$ should
include at least the information on the state $\rho _{n}$ or the unitary
operation $U_{n}$ based on different aims, since one hopes to extract the
information by the control qubit. That is, $L(\rho _{f})$ should not vanish.
In this case, we can obtain the similar theorem as Theorem 1 and Theorem 2.

\textbf{Theorem 3.} Let the entangling power of the DQC1 circuit subject to $%
\rho _{n}$ be $E_{p}^{P_{3}}\left( \tilde{U}_{n},\rho _{n}\right) $, $%
E_{p}^{P_{3}}\left( \tilde{U}_{n},\rho _{n}\right) $ does not vanish for $%
L(\rho _{f})>0$; 
\begin{equation}
1-\text{Tr}\sqrt{U_{n}\rho _{n}U_{n}^{\dagger }\rho _{n}}\leq
E_{p}^{P_{3}}\left( \tilde{U}_{n},\rho _{n}\right) \leq \sqrt{1-\left\vert 
\text{Tr}U_{n}\rho _{n}\right\vert ^{2}},
\end{equation}%
and 
\begin{equation}
E_{p}^{\mathbf{P}}\left( \tilde{U}_{n},\rho _{n}\right) =(\lambda
_{1}-\lambda _{2})E_{p}^{P_{3}}\left( \tilde{U}_{n},\rho _{n}\right) ,
\end{equation}%
where the superscript $P_{3}$ means $\mathbf{P}=(0,0,1)^{T}$, the
superscript $\mathbf{P}$ is the general case in $\rho _{c}$ and $\lambda
_{i} $ are the square root of the eigenvalues of the matrix $\rho _{c}\sigma
_{z}\rho _{c}^{\ast }\sigma _{z}$ in decreasing order.

\textbf{Proof.} The proof is quite similar to that of Theorem 2. The details
are given in the Appendix. One can easily check that Theorem 1 and Theorem 2
are covered in this theorem. \hfill $\blacksquare $

\section{Conclusions}

In summary, we have shown that the normalized trace of the unitary
transformation $U_{n}$ can be directly described by the entangling power of
the circuit of the DQC1. In addition, the entangling power also determines
the intrinsic complexity of the evaluation of the normalized trace of the
measured unitary matrix. In this sense, we think that the entangling power
could be the signature of the quantum advantage of the DQC1.  Furthermore,
we also present a generalization of the current DQC1 circuit to other QIPTs.
We find that the entangling power will not vanish if the information can be
extracted by the control qubit.

\section{Acknowledgements}

Yu thanks V. Vedral for his valuable suggestions and comments and thanks
Mile Gu for the discussions with him. This work was supported by the
National Natural Science Foundation of China, under Grant No. 11175033 and
`973' program No. 2010CB922904 and the Fundamental Research Funds of the
Central Universities, under Grant No. DUT12LK42.

\section{Appendix: Brief proof of Theorem 3}

For the general state $\rho _{n}$, it is difficult to calculate the exact
entangling power. Here we give its bounds and show that the entangling power
is always present, if one can extract some information. The second
inequality in Eq. (30) is obviously correct, which is similar to Eq. (8).
Now we give the brief proof for the first inequality in Eq. (30). According
to the definition of our entangling power, we have, for $\rho
_{n}=\sum\limits_{i}q_{i}^{\prime }\left\vert \varphi _{i}^{\prime
}\right\rangle \left\langle \varphi _{i}^{\prime }\right\vert $, 
\begin{eqnarray}
E_{p}^{P_{3}}\left( \tilde{U}_{n},\rho _{n}\right)  &=&\max_{\left\{
q_{i},\left\vert \varphi _{i}\right\rangle \right\}
}\sum\limits_{i}q_{i}^{\prime }\sqrt{2\left( 1-Tr\left( \rho _{r}^{i}\right)
^{2}\right) }  \notag \\
&\geq &1-\max_{\left\{ q_{i},\left\vert \varphi _{i}\right\rangle \right\}
}\sum\limits_{i}q_{i}^{\prime }\left\vert \left\langle \varphi _{i}^{\prime
}\right\vert U_{n}\left\vert \varphi _{i}^{\prime }\right\rangle \right\vert 
\notag \\
&=&1-\max_{\tilde{T}}\sum\limits_{i}\left\vert \tilde{T}^{\dagger }\sqrt{%
\tilde{M}}\tilde{\Phi}^{\dag }U_{n}\tilde{\Phi}\sqrt{\tilde{M}}\tilde{T}%
\right\vert _{ii}  \notag \\
&=&1-\text{Tr}\sqrt{U_{n}\rho _{n}U_{n}^{\dagger }\rho _{n}},
\end{eqnarray}%
where the first inequality holds because the purity of $\rho _{r}^{i}$ is
not more than $1$, and we consider the relation between the
eigendecomposition $\rho _{n}=\tilde{\Phi}\tilde{M}\tilde{\Phi}^{\dag }$ and
the other decompostions similar to those between Eqs. (15) and (16). One can
find that the lower bound of $E_{p}^{P_{3}}\left( \tilde{U}_{n},\rho
_{n}\right) $ given in Eq. (30) will vanish if $[U_{n}$, $\rho _{n}]=0$.
However, in this case, one can easily prove that $E_{p}^{P_{3}}\left( \tilde{%
U}_{n},\rho _{n}\right) $ will not vanish unless $U_{n}=e^{i\theta }\mathbf{1%
}_{n}$ which leads to zero $L(\rho _{f})$. Thus we show that any non-trivial
QIPT with DQC1 circuit is accompanied with entangling power.

In order to show that Eq. (31) holds, we have to rewrite the initial control
qubit $\rho _{c}$ given in Eq. (28). Similar to Eqs. (14,15), any
decomposition of $\rho _{c}$ can be related to its eigendecomposition $\rho
_{c}=\Phi ^{\prime }M^{\prime }\Phi ^{\prime \dag },$ where 
\begin{eqnarray}
\Phi ^{\prime } &=&\left( 
\begin{array}{cc}
\cos \frac{\theta }{2} & \sin \frac{\theta }{2} \\ 
e^{i\phi }\sin \frac{\theta }{2} & -e^{i\phi }\cos \frac{\theta }{2}%
\end{array}%
\right) , \\
M^{\prime } &=&\frac{1}{2}\left( 
\begin{array}{cc}
1+\frac{1}{\Gamma } & 0 \\ 
0 & 1-\frac{1}{\Gamma }%
\end{array}%
\right)
\end{eqnarray}%
with $\cos \theta =P_{3}/\Gamma $, $\Gamma =\sqrt{\sum%
\limits_{k=1}^{3}P_{k}^{2}}$. Thus any decomposition of $\rho _{c}=\Psi
^{\prime }W^{\prime }\Psi ^{\prime \dag }$ can be written by $\Psi ^{\prime }%
\sqrt{W^{\prime }}=\Phi ^{\prime }\sqrt{M^{\prime }}T^{\prime }$ with $%
T^{\prime }$ the right unitary matrix. Substitute any one possible pure
state $[\Psi _{11}^{\prime },\Psi _{12}^{\prime },\cdots ]^{T}$ and $\rho
_{n}$ into the DQC1 circuit, one will arrive at the final state as%
\begin{equation}
\left\vert \gamma _{j}^{\prime }\right\rangle =\frac{1}{\sqrt{r_{j}^{\prime }%
}}\left( x_{j}^{\prime }\left\vert 0\right\rangle \left\vert \varphi
_{i}^{\prime }\right\rangle +y_{j}^{\prime }\left\vert 1\right\rangle
U_{n}\left\vert \varphi _{i}^{\prime }\right\rangle \right) ,
\end{equation}%
where 
\begin{eqnarray}
x_{j}^{\prime } &=&\frac{1}{\sqrt{2}}\left( T_{1j}^{\prime }\left( \Phi
_{11}^{\prime }+\Phi _{21}^{\prime }\right) \sqrt{1+\frac{1}{\Gamma }}\right.
\notag \\
&&+\left. T_{2j}^{\prime }\left( \Phi _{12}^{\prime }+\Phi _{22}^{\prime
}\right) \sqrt{1-\frac{1}{\Gamma }}\right) , \\
y_{j}^{\prime } &=&\frac{1}{\sqrt{2}}\left( T_{1j}^{\prime }\left( \Phi
_{11}^{\prime }-\Phi _{21}^{\prime }\right) \sqrt{1+\frac{1}{\Gamma }}\right.
\notag \\
&&+\left. T_{2j}^{\prime }\left( \Phi _{11}^{\prime }-\Phi _{21}^{\prime
}\right) \sqrt{1-\frac{1}{\Gamma }}\right)
\end{eqnarray}%
and $r_{j}^{\prime }=\left\vert x_{j}^{\prime }\right\vert ^{2}+\left\vert
y_{j}^{\prime }\right\vert ^{2}$ with $\left\vert T_{1j}^{\prime
}\right\vert ^{2}+\left\vert T_{2j}^{\prime }\right\vert ^{2}$ $\leq 1$ and 
\begin{equation}
\sum_{j}\left\vert T_{1j}^{\prime }\right\vert ^{2}=\sum_{j}\left\vert
T_{2j}^{\prime }\right\vert ^{2}=1.
\end{equation}%
Similar to Eq. (20), the entangling power can be given by%
\begin{eqnarray}
E_{p}^{\mathbf{P}}\left( \tilde{U}_{n},\rho _{n}\right) &=&\min_{\left\{
r_{j}^{\prime },\left\vert \gamma _{j}^{\prime }\right\rangle \right\}
}\max_{\left\{ \tilde{q}_{i},\left\vert \tilde{\varphi}_{i}\right\rangle
\right\} }\sum\limits_{i,j}\sqrt{2}\tilde{q}_{i}\left( r_{j}^{\prime
2}\right.  \notag \\
&&-\left\vert x_{j}^{\prime }\right\vert ^{4}-\left\vert y_{j}^{\prime
}\right\vert ^{4}\left. -2\left\vert x_{j}^{\prime }\right\vert
^{2}\left\vert y_{j}^{\prime }\right\vert ^{2}\left\vert \left\langle \tilde{%
\varphi}_{i}\right\vert U_{n}\left\vert \tilde{\varphi}_{i}\right\rangle
\right\vert ^{2}\right) ^{\frac{1}{2}}  \notag \\
&=&\min_{\left\{ r_{j},\left\vert \gamma _{j}\right\rangle \right\}
}\sum\limits_{j}2\left\vert x_{j}^{\prime }\right\vert \left\vert
y_{j}^{\prime }\right\vert  \notag \\
&&\times \max_{\left\{ q_{i},\left\vert \varphi _{i}\right\rangle \right\}
}\sum\limits_{i}\tilde{q}_{i}\sqrt{1-\left\vert \left\langle \tilde{\varphi}%
_{i}\right\vert U_{n}\left\vert \tilde{\varphi}_{i}\right\rangle \right\vert
^{2}}  \notag \\
&=&\min_{\left\{ r_{j},\left\vert \gamma _{j}\right\rangle \right\}
}\sum\limits_{j}2\left\vert x_{j}^{\prime }\right\vert \left\vert
y_{j}^{\prime }\right\vert E_{p}^{P_{3}}\left( \tilde{U}_{n},\rho
_{n}\right) .
\end{eqnarray}%
Consider Eqs. (36,37), one will have 
\begin{eqnarray*}
\min_{\left\{ r_{j},\left\vert \gamma _{j}\right\rangle \right\}
}\sum\limits_{j}2\left\vert x_{j}^{\prime }\right\vert \left\vert
y_{j}^{\prime }\right\vert &=&\min_{\left\{ r_{j},\left\vert \gamma
_{j}\right\rangle \right\} }\sum\limits_{j}\left\vert T^{T}M\Phi ^{T}\sigma
_{z}\Phi MT\right\vert _{jj} \\
&=&\lambda _{1}-\lambda _{2},
\end{eqnarray*}%
where $\lambda _{i}$ are the square root of the eigenvalues of the matrix $%
\rho _{c}\sigma _{z}\rho _{c}^{\ast }\sigma _{z}$ in decreasing order. The
proof is completed.

\end{document}